# Geoneutrinos from the rock overburden at SNO+

V. Strati[1,2], S. A. Wipperfurth[3], M. Baldoncini[2,4], W. F. McDonough[3,5], S. Gizzi[2] and F. Mantovani[2,4].

[1]INFN, Legnaro National Laboratories, Legnaro, Italy
[2]University of Ferrara, Department of Physics and Earth Sciences, Ferrara, Italy
[3]Department of Geology, University of Maryland, College Park, Maryland, USA
[4]INFN, Ferrara Section, Ferrara, Italy
[5]Department of Earth and Planetary Materials Science and Research Center for Neutrino Science, Graduate School of Science, Tohoku University, Sendai, Miyagi 980-8578, Japan.

E-mail: strati@fe.infn.it

**Abstract**. SNOLAB is one of the deepest underground laboratories in the world with an overburden of 2092 m. The SNO+ detector is designed to achieve several fundamental physics goals as a low-background experiment, particularly measuring the Earth's geoneutrino flux. Here we evaluate the effect of the 2 km overburden on the predicted crustal geoneutrino signal at SNO+. A refined 3D model of the 50 × 50 km upper crust surrounding the detector and a full calculation of survival probability are used to model the U and Th geoneutrino signal. Comparing this signal with that obtained by placing SNO+ at sea level, we highlight a $1.4^{+1.8}_{-0.9}$ TNU signal difference, corresponding to the ~5% of the total crustal contribution. Finally, the impact of the additional crust extending from sea level up to ~300 m was estimated.

## 1. Introduction

SNO+ is a multipurpose kiloton-scale liquid scintillation detector located at (2092 ± 6) m underground at SNOLAB [1], in the heart of Vale's Creighton mine close to Sudbury (Canada). SNO+ is designed to address a variety of physics goals in the low energy neutrino sector, including the study of geoneutrinos [2]. These electron antineutrinos are emitted in beta minus decays occurring along the $^{238}$U and $^{232}$Th decay chains and offer a unique direct probe of the composition of the Earth's interior, possibly allowing for a discrimination among different Bulk Silicate Earth compositional models [3] [4] [5].

Gathering some insights into the mantle contribution to the geoneutrino signal at SNO+ can be conceivably pursued providing a detailed understanding of the CANDU dominated reactor background [6] and a refined regional-scale model of the Close Upper Crust (CUC), i.e. the 50 × 50 km upper crust surrounding the detector [7]. Located under a flat overburden, SNO+ is sited on the Canadian Shield at the boundary of the Superior and Grenville Provinces; the detector is enclosed in heterogeneous and complex lithologies that resulted from a 1.85 Ga meteorite impact event. This work addresses the question "how does the in-depth location of the SNO+ detector affect the prediction on the geoneutrino signal from the CUC?". Although the argument could seem self-explanatory in terms of geometrical effects, the latter have to be convolved with signal variations associated to the different geochemical and geophysical features of the surrounding geological units.

The geoneutrino signal from the CUC was evaluated considering SNO+ at its present position and compared to a signal predicted by placing the detector at sea-level, and thirdly, by detailing the contribution from the ~300 m thick rock that sits above sea-level (Figure 1).

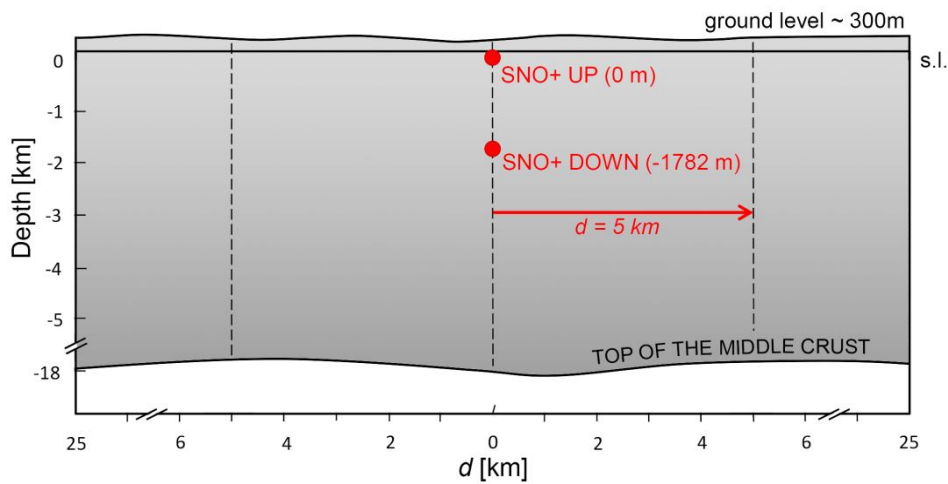

Figure 1 – Schematic representation of the SNO+ detector locations adopted in this study and of the upper crust surrounding SNOLAB. The SNO+ *up position* and *down position* correspond to a depth equal to the sea level (s.l.) and to 1782 m below sea-level, respectively. The 50 × 50 km Close Upper Crust (CUC) has an average thickness of 18.4 km and the topography of the region has a typical elevation of ~300 m above sea-level. The distance $d$ is the radius of the cylinder drawn around the detector position corresponding to the X axis of plots in Figure 3.

## 2. A 3D model integrating geological, geophysical and geochemical data

When calculating the flux of electron antineutrinos from above the detector, one consideration is an extraterrestrial source. Aside from short flux bursts from astrophysical objects outside of the solar system, two bodies can be considered as possible extraterrestrial sources: the Sun and the Moon. A simple model for both has the Sun at about a million times the mass of the Earth and standing at $1.5 \times 10^{11}$ meters separation distance, while the Moon is $3.8 \times 10^{8}$ meters away and 1.2% of the Earth's mass. Thus, the U and Th masses would be $M(U)_{Moon}$ ~$1 \times 10^{15}$ kg and $M(Th)_{Moon}$ ~$5 \times 10^{15}$ kg for the Moon and $M(U)_{Sun}$ ~$8 \times 10^{17}$ kg and $M(Th)_{Sun}$ ~$3 \times 10^{18}$ kg for the Sun. The expected oscillated flux from the Sun and Moon would be ~$2 \times 10^{-2}$ and ~4 electron antineutrinos/s/cm$^2$, respectively.

Therefore, these flux contributions are considered negligible compared with a typical Earth's flux of ~$10^6$ electron antineutrinos/s/cm$^2$.

The prediction of the geoneutrino signal is based on the numerical 3D geological model of the CUC presented in [7] in which nine representative aggregate units of exposed lithologies are geologically characterized, geophysically constrained, and geochemically defined using new analyses and compiled databases. The geophysical and geochemical properties of the units adopted in this study are summarized in Table 3 of [7].

The geological units were simplified using the published 1:250,000 scale Bedrock Geology of Ontario [8] and supplemented by available subsurface data. The crustal structures of the 9 units were defined by combining multiple inputs: (i) the contacts of the simplified geological map, (ii) the digital elevation model produced by the Shuttle Radar Topographic Mission (SRTM) [9], (iii) the map of depth of the top of the middle crust presented in [10], (iv) the 2.5 D geological models along six profiles used for the construction of the 3D model reported in [11] and (v) five virtual cross sections derived from the model developed in [10]. Figure 2 provides a 3D model visualization accompanied by two orthogonal sections (NS and EW). The mine and the SNO+ detector are placed in the contact zone between two geological units, the Huronian Supergroup and minor Intrusions (HI) and Norite-Gabbro (NG) unit. The latter, intruding the HI, is part of the main mass of the Sudbury Igneous Complex, an impact melt sheet originated by a meteorite impact (1.85 Ga) and subsequently filled by sediments of the Whitewater Group (Figure 2).

We collected and analyzed by HPGe gamma-ray spectroscopy 109 rock samples for their U and Th abundances from representative units in the CUC area. A subset of samples with low Th and U abundances were also analyzed by ICPMS [7]. The reference geological map set the rationale for the sampling campaign; the number of samples collected scaled as a function unit exposure area, its estimated volume, and its proximity to the detector. Globally the units constituting more than 90% of the volume were characterized with more than 30 samples, allowing for a statistical study of the frequency distributions and for applying the Kolmogorov-Smirnov statistical test to probe the normal and lognormal tendencies of the U and Th distributions. At the same time, the adopted method provided the base for a robust treatment of the geochemical uncertainties and for the study of the correlation between U and Th abundances via a bivariate analysis. These issues have a strong impact on the geoneutrino signal estimations and for this reason cannot be disregarded in the signal calculation and in turn in the evaluation of its uncertainty.

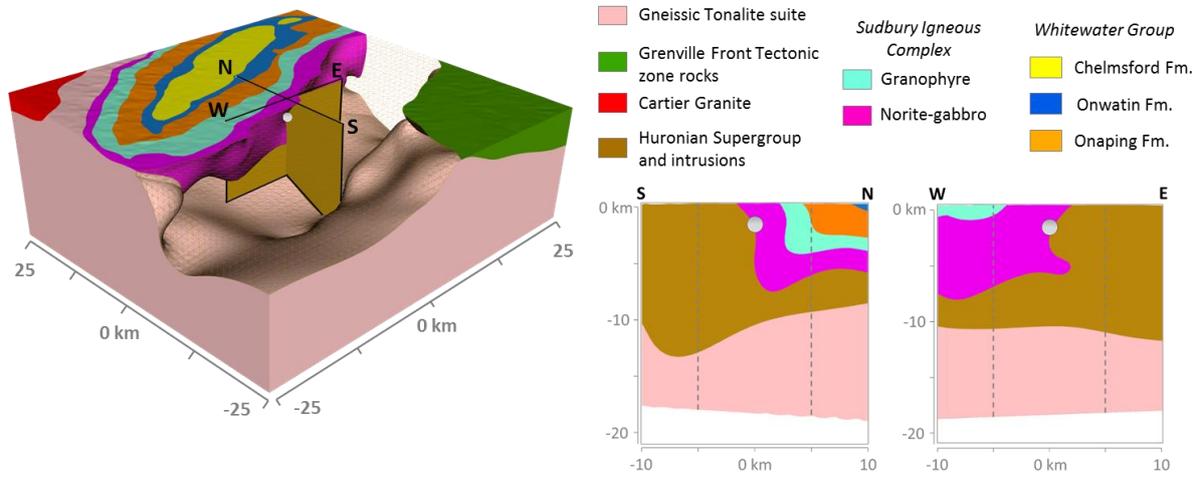

Figure 2 – A 3D view of the model with the HI unit removed in order to visualize the in-depth location of the SNO+ detector (light grey dot). On the right is reported the visualization of the model along two orthogonal 2D sections horizontally centred in SNO+. The detector is located in the contact zone between NG and HI units. Dashed grey line corresponds to the radius $d$ of the cylinder drawn around the detector position (see Figure 1).

## 3. Geoneutrino signal analysis

The 3D model described in Section 2 was discretized in voxels of 0.1×0.1×0.1 km dimensions to which specific values of density and radioisotope abundances were assigned, which are necessary for assessing the distinct U and Th geoneutrino activities, i.e. the average number of antineutrinos produced by individual voxels per unit time. By weighting each activity with the corresponding geoneutrino spectrum [4] and by scaling for the $1/4\pi r^2$ spherical factor and for the three flavor survival probability [13] with up to date mixing parameters [13], the oscillated geoneutrino flux at SNO+ is estimated. Geoneutrino fluxes are converted to signals in TNU (Terrestrial Neutrino Unit: corresponds to one geoneutrino event per $10^{32}$ free target protons per year) by accounting for the Inverse Beta Decay cross section [14].

The effect of the SNO+ burial was investigated by computing the expected geoneutrino signal for two positions having depths of 0 m a.s.l (*up position*) and 1782 m b.s.l. (*down position*), with the latter corresponding to the actual SNOLAB location (see Figure 1). In Figure 3 the cumulative signal produced separately by the HI and NG units and by the entire upper crust is reported for both SNO+ locations as function of the radius of the cylinder $d$ (see Figure 1). The choice of the cylindrical symmetry allows direct comparison of the geoneutrino signals predicted at the *up position* and *down position* since they are produced by identical volumetric sources. Thus, it is possible to jointly test the combined effect of a geometrical vertical translation and a displaced immersion of the SNO+ detector in the local geological structures.

For each component the cumulative signal at $d$ = 5 km for the *down position* exceeds the corresponding one predicted for the *up position*, as one would expected simply in terms of the flux spherical scaling factor. However, a peculiar feature is that the NG and HI curves respectively do and do not intersect for the *up position* and *down position*, which is also related to a different radial starting point of the HI curve (Figure 3). Indeed, although for the *up position* the HI unit is on average more distant to the detector with respect to the *down position*, the higher U and Th contents make the HI curve pass over the NG curve at ~3 km radial distance.

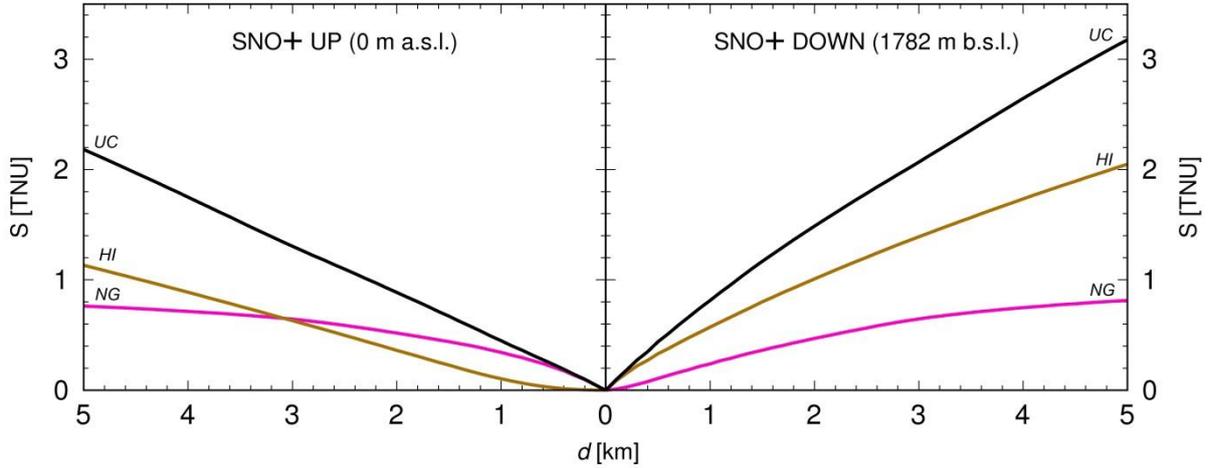

Figure 3 - Right and left panels show the cumulative geoneutrino signal for SNO+ in the *down position* (1782 b.s.l.) and *up position* (0 m a.s.l.), respectively, as function of the radial distance *d* from the cylinder vertical axis (see Figure 1). The magenta, brown and black curves refer to the signal generated by the NG and HI geological units and by the entire upper crust (UC), respectively.

The presence of the thick overburden is also globally quantified in terms of predicted geoneutrino signal in the CUC region for the *up position* and *down position*. Central values and asymmetric standard deviations are estimated by combining their results with a Monte Carlo simulation that accounts for the individual contributions produced by the 9 CUC geological units. Each unit is characterized by its proper geophysical and geochemical uncertainties [7]. For SNO+, the expected signal from the CUC in the *up position* is $S_{UP}= 6.3^{+5.6}_{-2.3}$ TNU, in the *down position* is $S_{DOWN} = 7.7^{+7.7}_{-3.0}$ TNU, while the contribution of the nearly flat, ~300 thick cover extending from sea level up is $S_{COVER}= 0.23^{+0.17}_{-0.06}$ TNU.

## 4. Conclusions
The 2092 m of overburden above the SNO+ detector is an efficient cosmic ray shield and also a significant source of geoneutrinos. For the first time, the effect of the in-depth location in the estimation of the geoneutrino signal is evaluated.

At the spatial scale of the CUC (i.e. the 50 × 50 km upper crust centred at the SNO+ location), the absolute signal difference estimated on the base of the refined 3D model [7] by placing the SNO+ detector at 1782 m b.s.l. and at 0 m a.s.l. (Figure 1) is of $1.4^{+1.8}_{-0.9}$ TNU.

A difference of ~1 TNU between the two cases (detector at sea-level and 1782 m below sea-level) is observed for the crustal signal emitted from a 5 km radial distance cylinder (Figure 3). Two main geological units (NG and HI) exhibit different signal strengths depending on the adopted vertical position of the detector. This result reflects the complex geology of the region, both in terms of shape and composition of the geologic units. These results provide insights that better inform a coherent geophysical and geochemical investigation of the SNO+ surroundings, which is a fundamental step for improving models of its geoneutrino signal. Incorrect adoption of a shallow (e.g., at sea level) position for the SNO+ detector induces an 18% underestimation on the CUC signal ($7.7^{+7.7}_{-3.0}$ TNU) and a 5% reduction on the total crustal signal ($31.1^{+8.0}_{-4.5}$ TNU) at SNO+ [7]. Distinction between mantle and crustal contributions to the total is a critical challenge and thus this point has a non-negligible impact on the result. Similarly, overburden signals for all detectors need to be considered when conducting a global data fit of the mantle and crustal contributions [5].